\documentclass[12pt,dvips]{article}

\usepackage{amsmath,amssymb,exscale}
\usepackage{array,multicol}
\usepackage{afterpage,float,flafter}
\usepackage{epsfig,rotating,pifont}
\usepackage{cite}
\makeatletter
\@addtoreset{equation}{section}
\makeatother

\title{
\vspace*{-0.8cm}
\begin{flushright}
\end{flushright}
\vspace{1cm} 
\Large\textbf{Fat Gravitons, the Cosmological Constant \\
 and Sub-millimeter Tests}
\vspace*{.5cm}
\author{\large \textbf{
Raman Sundrum\footnote{email: ~ sundrum@pha.jhu.edu} 
}\\
\\
\emph{
Department of Physics and Astronomy} \\ 
\emph{Johns Hopkins University} \\ 
\emph{3400 North Charles St}. \\ 
\emph{Baltimore, MD 21218-2686}}}

\date{}
\begin{document}
\maketitle
\thispagestyle{empty}
\vspace*{.5cm}
  
\begin{abstract} 
We revisit  the proposal that the 
 resolution of the Cosmological Constant Problem involves 
a sub-millimeter 
 breakdown of the point-particle approximation for gravitons.
 No fundamental description of such a breakdown, which simultaneously preserves the 
point-particle nature of matter particles, is yet known.  
However, 
basic aspects of the self-consistency 
of the idea, such as preservation of the macroscopic Equivalence Principle
while satisfying quantum naturalness of the cosmological constant,  
are addressed in this paper within a Soft Graviton Effective Theory.
It builds on Weinberg's analysis of soft graviton couplings and 
standard heavy particle effective theory, and  minimally encompasses the experimental
regime of soft gravity coupled to hard matter.   
A qualitatively distinct  signature for short-distance tests of gravity is 
discussed, bounded by naturalness to appear above approximately 20 microns.

\end{abstract} 
  
\newpage 
\renewcommand{\thepage}{\arabic{page}} 
\setcounter{page}{1}


\section{Introduction}

Imagine an alien species, sophisticated enough to know the overarching 
principles of quantum mechanics and relativity, but whose particle physics
expertise (or funding) can only engineer or observe momentum transfers below
 $10^{-3}$ eV. While they have access to heavy, macroscopic sources, 
the only fundamental fields 
and particles they know are the metric of General Relativity (GR), soft
electromagnetic fields and perhaps some neutrinos.
The alien theoreticians have nevertheless 
synthesized the various tools of quantum field theory from 
the big principles. Superstring theory is also flourishing.
Phenomenologists have  
put in place a minimal effective field theory cut off by $10^{-3}$ eV, which 
accomodates the data below this scale while being agnostic about physics above.

The aliens  have also run into the Cosmological 
Constant Problem (CCP). (For a review see Ref. \cite{weinberg}.) Actually, 
since  the observed ``dark energy'' density of the cosmos is 
$\sim (10^{-3}$ eV)$^4$ \cite{ccexpt} \cite{pdg}, 
their minimal effective theory is not presently
fine-tuned. However, if 
new experiments above $10^{-3}$ eV continue to support the minimal effective 
theory, now with a larger UV cutoff, then the cosmological constant 
would be fine-tuned.   Naturalness therefore predicts new 
physics just above $10^{-3}$ eV, acting to cut off the 
quartic divergences in the cosmological constant within the effective theory.
 The aliens are 
therefore quite excited about new short distance tests of gravity, 
$< 0.1$ mm $\sim (10^{-3}$ eV)$^{-1}$, along with other ``high-energy'' experiments.
 They imagine that they might discover sub-millimeter strings cutting 
off all of point-particle 
effective field theory, or 
supersymmetry enforcing cancellations in radiative corrections 
to the cosmological constant. Or perhaps something no one has thought of.

We, on the other hand, seem less excited that experiments have the answer. 
We already know too much. Our 
particle physicists have probed momentum transfers  all the way up to a TeV 
without finding sub-millimeter supersymmetry or strings. Effective field 
theory of the Standard Model (SM) coupled to GR with  a TeV cutoff beautifully
accounts for all the data, but now
the cosmological constant is unavoidably fine-tuned. 
There is ``no-go'' theorem forbidding new light fields from relaxing the cosmological constant
\cite{weinberg}.
The door through which 
new sub-millimeter 
gravitational physics might enter into a solution of the CCP seems firmly
shut. 

The purpose of this paper is to pry open this door a little.
 An important first step is to notice that the TeV scale 
effective theory which leads to the CCP involves a tremendous 
extrapolation of standard GR to far shorter distances than 
gravity has been experimentally probed, in order to accomodate the wealth of 
SM data.\footnote{Such an extrapolation should certainly not be taken for granted, as 
dramatically illustrated by
 noting that present data cannot distinguish  a theory 
with a gravity-only large extra dimension with a size of order 
$0.1$ mm (similar to the well-known proposal of Ref. \cite{add} with two extra dimensions)
 from the usual 4D theory.} 
Naively, this observation has no bearing on the CCP, since SM
corrections to the IR effective cosmological constant reside in the gravitational 
effective action, $\Gamma_{eff}[g_{\mu \nu}]$, evaluated for extremely 
soft gravitational fields. The SM fields are hard and off-shell in general in such 
corrections, but then hard SM processes are well understood up to a TeV. 
Nevertheless, the central point of Ref. \cite{97} was to argue
 that (virtual) high energy
contributions to an effective 
action, $\Gamma_{eff}[A]$, of a sector, A, from integrating out a different sector, B, 
cannot be robustly determined (or even roughly estimated) 
without knowing the high energy dynamics and degrees of 
freedom of {\it both sectors, A and B}. This conclusion does not follow from 
standard Feynman diagram calculations, but rather by re-thinking whether certain diagrams 
are warranted at all. Ref. \cite{97} illustrated the general claim by 
studying an analog system   built out of QCD,
where sector A undergoes a radical, but well hidden, change in its degrees of freedom, 
from light pions at low energies to quarks and gluons at high energies. 
The tentative conclusion drawn for the real CCP was that a drastic change in 
gravitational degrees of freedom above $10^{-3}$ eV, 
akin  to compositeness\footnote{The graviton could not literally be a composite of 
a Poincare invariant quantum field theory, by the theorem of Ref. \cite{ww}. However, the 
physical manifestations of compositeness are compatible with the graviton, as string theory 
perfectly illustrates.}, could suppress high energy 
SM contributions to $\Gamma_{eff}[g_{\mu \nu}]$. If this is correct, then even though we know
much more than the aliens about the SM, we have every reason to be as excited about 
sub-millimeter tests of gravity \cite{gravexpt1} \cite{gravexpt2} 
where the new degrees of freedom would be revealed.

The present paper will further discuss the case for such 
a resolution to the CCP by a ``fat graviton'', and how it can be tested experimentally 
in the near future. We do not construct a fundamental theoretical model of 
fat gravitons and point-like matter here. Instead we pursue a more modest goal. We argue that 
such a resolution to the CCP, and the relevance therefore of short-distance gravity experiments, 
is not ruled out by general considerations and principles, despite the fact that these 
considerations seem  at first sight to strongly exclude any such scenario.

Many aspects of 
self-consistency are addressed by an effective field theory formalism we will call 
Soft Graviton Effective Theory (SGET). It blends together aspects of 
Weinberg's analysis of soft graviton couplings \cite{wsoft}
 with  standard heavy 
particle effective theory \cite{hpet}. It is hoped that 
the development of such an effective field theory description will make the ideas 
precise enough to pursue more fundamental model-building, say within string theory, or 
to identify and pursue phenomenological implications. On the other hand, 
the more precise description of fat gravity may lead to falsification, either by 
experimental means or by proving ``no-go'' theorems. At least we will know 
for sure then that the door is shut. 

The discussion of the CCP in this paper suffers from some significant limitations. 
There are issues 
related to the CCP which involve cosmological time evolution. The discussion here takes 
a rather static view of the problem, focussing on SM quantum corrections. A key diagnostic 
tool for any new mechanism for the CCP is to consider its behavior in cases where there are 
multiple (metastable) vacua. It is certainly very interesting 
to pursue these considerations in the 
case of the present proposal, but the result is not yet conclusive and a discussion is 
deferred for later presentation. There is undeniably a new scale in gravity provided by the
observed dark energy density \cite{ccexpt}. While in 
this paper it is related to the ``size'' of the fat graviton, this size 
is not predicted from other considerations but taken as input. 
Its constancy over cosmological times is also not determined.

The proposition that the small vacuum energy density might translate
 by naturalness into a 
sub-millimeter scale for new gravitational physics was made in Ref. \cite{banks} (although 
the primary subject of Ref. \cite{banks} is a quite different approach to the CCP).
The idea  that this new physics involves a sub-millimeter breakdown 
in  point-like gravity \cite{97} 
has been further discussed in the extra-dimensional proposal of Ref. \cite{gia}. 

The layout of this paper is as follows. Section 2 re-analyses the robustness of the CCP 
in standard effective field theory and how it rests on the presumption that the graviton 
is point-like and able to mediate hard momentum transfers.
 Section 3 considers the consequences of rejecting this presumption, that is, entertaining the 
possibility of a ``fat graviton''. We see that there is now a loop-hole in the CCP, but that 
the macroscopic consequences of GR and the Equivalence Principle are necessarily preserved. 
Section 4 discusses experimental/observational contraints and predictions following from 
a fat graviton resolution of the CCP. In particular, naturalness predicts a non-zero 
cosmological constant, now however set only by the graviton ``size''. There is rather a sharp 
prediction for where fat graviton modifications of Newton's Law should appear, and the 
qualitative form they take. In Section 5, we begin construction of a Soft Graviton Effective 
Theory (SGET) which satisfies basic principles, captures the physics of hard SM processes 
as well as soft graviton exchanges between SM matter, but {\it does not extrapolate 
standard GR to short distances}. 
 It is demonstrated that this effective theory, capable of 
minimally capturing our present experimental regimes, does not give robust contributions to the 
cosmological constant from heavy SM physics, thereby clarifying the fat-graviton 
loophole. Section 6 provides conclusions.

\section{Robustness of the CCP}

\begin{figure}[t]
\centering
\epsfig{file=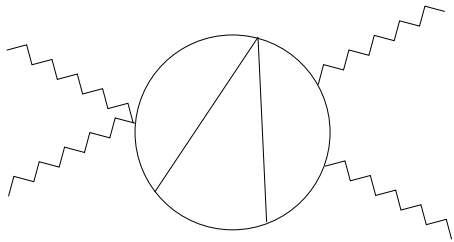,width=.35\linewidth}
\caption{Typical SM quantum contributions to $\Gamma_{eff}[g_{\mu \nu}]$. 
Jagged lines are gravitons and smooth lines are SM particles.}
\end{figure} 

The quantum contributions to the cosmological constant which dominate in standard effective 
field theory, and 
appear most robust, arise 
from Feynman diagrams such as Fig. 1, 
with SM matter loops and very soft graviton external lines. Diagrams with different 
numbers of external gravitons correspond to different terms in the expansion of the 
cosmological term about flat space, 
\begin{equation}
\label{sqrtg}
\sqrt{-g} = 1 + \frac{ h_{\mu}^{\mu}}{2 M_{Pl}}  + ..., ~ ~ 
g_{\mu \nu} \equiv \eta_{\mu \nu} + \frac{ h_{\mu \nu}   }{  M_{Pl}}. 
\end{equation}
Of course the tadpole term implies that flat space is destabilized by a cosmological 
constant, but as long as it is small the flat space diagrams still provide a convenient 
way to expand the leading effects.

Diagrams such as Fig. 1 make 
${\cal O}(\Lambda_{UV}^4/16 \pi^2)$ contributions to the effective cosmological constant,
when $\Lambda_{UV}$ is taken to be a typical general coordinate invariant (GCI) cutoff.
For $\Lambda_{UV} \geq$ TeV, the 
contribution is many orders of magnitude larger than the observed cosmological constant 
and the situation is highly unnatural. 
To search for the most robust contributions to the 
cosmological constant let us be more optimistic about the sensitivity to the 
true nature of 
the physics above a TeV which cuts off the diagrams (perhaps some type of stringiness). 
A simple way to 
do this is to calculate using dimensional regularization. However, this still results 
in finite contributions to the vacuum energy from known SM masses and interactions,
\begin{equation}
\label{smrobust}
\sim ~ \sum_{SM} \frac{(-1)^{F_{SM}}}{16 \pi^2} m_{SM}^4 \ln m_{SM} + 
{\cal O}(\frac{\lambda_{SM}}{(16 \pi^2)^2} m_{SM}^4), 
\end{equation}
which are still so large that the CCP is not 
much diminished. Here, the first term is due to just the free particle zero-point 
energies, while the second term is sensitive to SM couplings, $\lambda_{SM}$. 
These contributions of diagrams such as Fig. 1 
seem theoretically very robust. Afterall, the couplings of the graviton lines are being 
evaluated for soft momenta, precisely where we are most confident about their couplings
to SM matter given the experimental success of GR. We can 
therefore use Eq. (\ref{sqrtg}) to evaluate the diagrams with any number of graviton 
legs, the one with no legs being the simplest way to compute the cosmological constant
contribution of course.
The remainder of the calculation involves the propagation and soft and hard quantum 
interactions of SM particles. Again,
we have tested all this extensively in particle physics experiments up to 
a TeV. Yet it is the  purpose of this paper is to look for loop-holes in the apparent 
robustness of the contributions to the cosmological constant from known SM physics.

Let us digress here from the main thrust of this paper to briefly discuss another well-known 
approach to the CCP which naively avoids the robustness of the contributions, Eq. 
(\ref{smrobust}). In this approach, GCI is replaced as the guardian symmetry of massless 
gravitons by Special Coordinate Invariance (SCI) 
\cite{sci} (for a review see Ref. \cite{weinberg}), consisting of only coordinate 
transformations with unit Jacobian and metrics with $\sqrt{-g} = 1$. In this approach, 
the cosmological constant appears as an extra integration constant of the (quantum) equations 
of motion, rather than being dynamically determined. In this way the CCP becomes an issue 
of  {\it initial conditions} and does  not relate to quantum corrections or, 
 by naturalness, to any testable 
new gravitational physics (which would be a shame, but of course 
this is not an argument against 
the idea). Even with SCI, however there is a formal (and quite possibly physically relevant 
in a more fundamental description of gravity) 
 objective meaning to 
Eq. (\ref{smrobust}). It follows by thinking of 
all $m_{SM}$ as having their
 origins as the VEVs of some source external fields, 
which may even vary somewhat 
in different parts of the universe.
We can formally write  
\begin{equation}
m_{SM} = \langle \chi(x) \rangle.
\end{equation}
Then Eq. (\ref{smrobust}) does contribute to dark energy, and if $\chi$ varies in different 
parts of the universe so do these contributions. In such a setting, the extra integration 
constant of SCI remains an exact spacetime constant which  
adds to the theoretically distinct and robust effects of Eq. (\ref{smrobust}). 
From this perspective, there  is not much difference between the implications of 
GCI and SCI, except that the ability in effective field theory to simply add an
arbitrary 
cosmological constant counterterm to any particular model is replaced by the ability to 
add the effect of an arbitrary integration constant. The need to fine-tune away quantum 
corrections from Eq. (\ref{smrobust}) remains intact, although with SCI this is done 
using the integration constant. From now on we return to taking GCI as the symmetry 
protecting the massless graviton.

\begin{figure}[t]
\centering
\epsfig{file=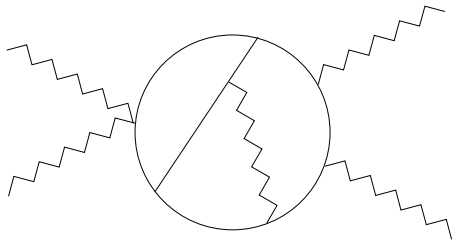,width=.35\linewidth}
\caption{Typical contribution to $\Gamma_{eff}$ involving SM as well as 
quantum gravity corrections.}
\end{figure} 

There are also subleading contributions (for $\Lambda_{UV} < M_{Pl}$) from diagrams 
such as Fig. 2, with graviton lines in the quantum loops. These quantum 
gravity contributions to the 
cosmological constant of ${\cal O}(\Lambda_{UV}^6/(16 \pi^2 M_{Pl}^2))$ are still 
significant from the point of view of naturalness (for $\Lambda_{UV} \geq$ TeV)
but one might hope that our imperfect understanding of quantum gravity makes these
contributions a less robust problem than diagrams such as Fig. 1. Further, they are 
certainly Planck-suppressed, and it is consistent for us to first neglect these 
effects and tackle instead the 
leading quantum corrections. In this paper, for simplicity
we will neglect quantum gravity corrections all together, deferring a treatment of this 
topic for later presentation. 

Let us return to consider the dominant contributions coming from purely SM loops. 
When we allow general graviton momenta,  Fig. 1 is a contribution to the gravitational 
effective action, $\Gamma_{eff}[g_{\mu \nu}]$, 
not just the cosmological constant. Let us ask, since the diagram is 
the cause of such concern, why we bother to include its contribution to $\Gamma_{eff}[g_{\mu \nu}]$ 
at all. A first response is that we are simply following the Feynman rules,
but let us inquire more deeply what fundamental principles are at stake if we simply throw out these
diagrams, but not the (well-tested) loop diagrams contributing to SM processes. Three principles 
stand out. 

(I) Unitarity: 
Fig. 1 unitarizes lower-order tree and loop processes of the form gravitons 
$\rightarrow$ SM ($+$ gravitons). That is, 
Fig. 1 has imaginary parts for general momenta 
following from unitarity and these lower order processes.
We have not yet seen such processes experimentally. Furthermore, when there are 
massive SM particles in the loops, the imaginary parts only exist once the 
external graviton momenta are above the SM thresholds. These are momenta for which 
quite generally we have not probed gravity (except that we know it is still so weak as 
to be invisible in experiments). If gravity is radically modified below such SM 
thresholds then we would have to radically modify diagrams such as Fig. 1.

(II) GCI: 
There are diagrams without imaginary parts in any momentum regime, but which  are 
required when we include diagrams with imaginary parts so as 
to maintain the GCI Ward identities, ultimately needed to 
protect theories of massless spin-2 particles. 
Note however that throwing out {\it all}
SM contributions to $\Gamma_{eff}[g_{\mu \nu}]$ is a perfectly GCI thing to 
do.

\begin{figure}[t]
\centering
\epsfig{file=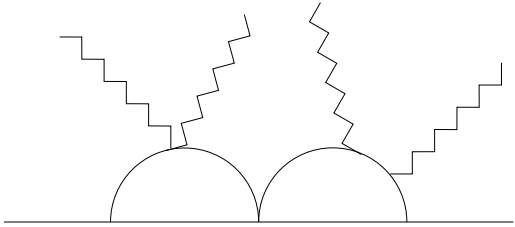,width=.4\linewidth}
\caption{Soft gravitons coupled to loop-correction to SM self-energy.}
\end{figure}

(III) Locality: In standard effective field theory one also has  non-vacuum 
SM diagrams with soft gravitons attached, such as Fig. 3. where soft gravitons 
couple to, and thereby measure,  loop corrections to a SM self-energy. We 
certainly do not want to throw this away since these 
contributions are absolutely crucial in maintaining the precisely tested equivalence 
of gravitational and inertial masses of SM particles and their composites.

Now, when Figs. 3 and 1 are  
viewed as position space Feynman diagrams (or better yet as first quantized sums over 
particle histories) it is clear that they are {\it indistinguishable locally in 
spacetime}, only globally can we make out their topological difference. Locality of 
couplings of the point particles in the diagrams does not allow us to contemplate 
throwing away Fig. 1 which we do not want, while retaining Fig. 3 which we need. 
The gravitons cannot take a global view of which diagram they are entering into before 
``deciding'' whether to couple or not. 
 Thus our earlier argument for the robustness of the CCP
 hinges on locality. We can dissect diagrams contributing to 
$\Gamma_{eff}[g_{\mu \nu}]$
 into small 
spacetime windows, and all the ingredient windows are 
well tested in other physical processes, albeit in globally different ways. 
We might contemplate dispensing with locality, 
but it seems to be the only way we explicitly know to have point-particle dynamics
 in a relativistic quantum
setting. However, see Refs. \cite{moffat} for an approach to the CCP with sub-millimeter 
non-locality, as well as 
Refs. \cite{banks} and 
\cite{savas} for extremely non-local approaches to the CCP.

\section{Room for a Fat Graviton} 

\subsection{Basic Notions}

\begin{figure}[t]
\centering
\epsfig{file=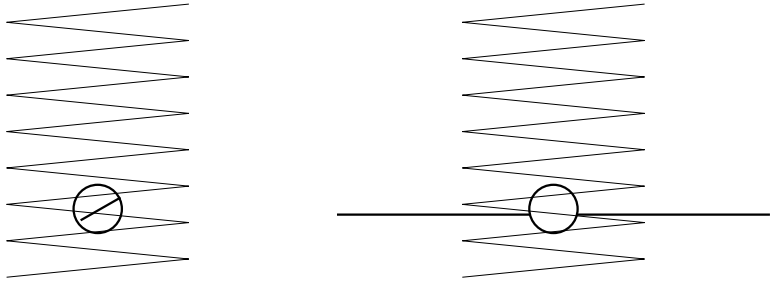,width=.6\linewidth}
\caption{A fat graviton can take a more global view of massive SM loop 
processes.}
\end{figure}

If some particles appearing  in the Feynman diagrams are secretly extended states 
then the constraints of locality, and the 
consequent robustness of the CCP, are 
weakened. Since we have tested the point-like nature of SM particles to very short 
distances, the only candidate for an extended state is the graviton itself. Indeed,
direct 
probes of the ``size'' of the graviton only bound it to be smaller than $0.2$ mm, 
following from short-distance tests of Newton's Law \cite{gravexpt1}. 
Let us grant the graviton a 
size, 
\begin{equation}
\ell_{grav} \equiv 1/\Lambda_{grav}.
\end{equation}
Such a ``fat graviton'' does not have to couple with point-like locality to SM loops, 
but rather with locality up to $\ell_{grav}$. In particular for $m_{SM} \gg 
\Lambda_{grav}$ a fat graviton can couple to SM loops {\it globally}, thereby 
evading reason (III) for the robustness of the CCP. 
To see this, note that locality up to $\ell_{grav}$ corresponds to standard locality of 
SM loops when only graviton wavelengths $> \ell_{grav}$ are allowed. Of course, for 
$m_{SM} \gg \Lambda_{grav}$ and graviton momenta $< \Lambda_{grav}$, diagrams such as 
Fig. 1 can be expanded as a series in the external momenta, that is a set of local 
vertices in spacetime. A cartoon of all this is given in Fig. 4.

Thus a theory with a fat graviton could distinguish between Figs. 3 and 1, possibly 
suppressing Fig. 1 
 while retaining Fig. 3 needed for the Equivalence Principle. It is at least 
conceivable. 

One naive objection is that among the diagrams contributing to the cosmological constant
is the one with no graviton external legs, corresponding to the first term on the 
right-hand side of Eq. (\ref{sqrtg}), that is, pure SM vacuum energy. Since the 
graviton does not appear in the diagram the size of the graviton appears irrelevant and 
incapable of suppressing the contribution. However, physically the cosmological term is a
self-interaction of the graviton field (defined about flat space say). Once we 
trust point-like diagrams we can use GCI to relate all of them for soft gravitons to
the diagrams with no gravitons and then it becomes a mathematical convenience to 
compute this latter class of diagrams. They do not have any direct physical significance
except as a short-hand for the diagrams with gravitons interacting. If the diagrams 
with graviton external lines are modified because gravitons are fat, there is no 
meaning to the diagram with no external lines. 
Indeed, notice that there is no 
physical consequence in an effective  lagrangian if in addition to 
a cosmological constant multiplying $\sqrt{-g}$, we add a pure 
constant term with no gravitational field attached. When we expand about flat space
 the extra constant modifies the first term in Eq. (\ref{sqrtg}). This shows that the 
cosmological term only has physical importance as a graviton coupling, and the 
size of the graviton is most certainly relevant to how SM loops can induce it.

Let us now return to the issue (I)  of the need to unitarize processes, 
gravitons $\rightarrow$ SM states ($+$ gravitons). We can most easily deal with this 
by making a precise assumption for what is an intuitive property of fat objects. 
We assume that hard momentum 
transfers $\gg \Lambda_{grav}$ via gravitational interactions are essentially forbidden,
that is they are extremely highly suppressed even beyond the usual Planck suppressions
of gravitational interactions. In particular for $m_{SM} \gg \Lambda_{grav}$,  
gravitons $\rightarrow$ SM states ($+$ gravitons) is suppressed, and the related 
unitarizing SM loop contributions, as well as GCI-related diagrams, are not required.

Throwing out all massive SM loop contributions 
to $\Gamma_{eff}[g_{\mu \nu}]$  is entirely consistent with the GCI of the 
soft graviton effective field theory below $\Lambda_{grav}$.

\subsection{ Naturalness of the Equivalence Principle}

While we have seen that the  general considerations (I -- III) for the 
robustness of massive SM loop contributions to the cosmological constant, and indeed
the whole of $\Gamma_{eff}[g_{\mu \nu}]$, are evaded by a fat graviton in principle, 
one can ask whether it requires  fine tunings even more terrible than the original 
CCP in order to understand why the fat graviton manages to 
couple to self-energies of SM states in the precise way to  maintain the 
equivalence between gravitational and inertial mass. After all, these self-energies,
for example the mass of a proton or of hydrogen, are determined by  short-distance
physics $\ll \ell_{grav}$, 
 which unlike a pointlike graviton, a fat graviton cannot probe.
In fact we shall see that there is really no option but that soft fat gravitons couple 
according to the dictates of the Equivalence Principle. The only miracle is that 
the fat graviton has a mode which is massless with spin 2 and couples somehow to matter.
 The only way for soft massless spin-2 particles to consistently couple is under the 
 protection of GCI \cite{protect}, which 
in turn leads to the Equivalence Principle macroscopically.
 The energy and momentum of SM states
 are macroscopic features which the fat graviton could 
imaginably couple to. They are determined microscopically but are measurable 
macroscopically, just as one can measure the mass density of a chunk of lead 
macroscopically, even though this density has its origins in and is sensitive to 
microscopic physics. In fact in Section 5 
we will show how  the leading couplings of gravity 
to SM masses and interaction energies can be recovered as a consequence only of 
GCI below $\Lambda_{grav}$, forced on us once we accept that our fat object 
contains an interacting massless spin-2 mode.

\subsection{ Soft and Hard SM effects}

Until now, we have been careless of the complication that the SM contains particles 
which can be lighter than $\Lambda_{grav}$, such as the photon, as well as particles 
which are heavier. Even with the fat graviton we cannot throw away loops of these 
lighter SM fields contributing to $\Gamma_{eff}[g_{\mu \nu}]$ 
because the soft components of these fields 
 are part of a standard effective field theory, including GR
 below $\Lambda_{grav}$ (that is, the effective theory known to the aliens). 
Soft gravitons can therefore scatter into soft 
electromagnetic radiation, so there are unitarity-required loops in the gravitational 
effective action. Indeed, we know that soft massless SM loops are not local on the 
scale $\Lambda_{grav}$, that is such loops are not expandable as local vertices for 
soft graviton momenta. Therefore, even a fat graviton cannot couple globally to such 
loops and the arguments for their robustness from locality now apply. A further 
complication is that the soft SM particles can couple to hard or massive SM particles.

\begin{figure}[t]
\centering
\epsfig{file=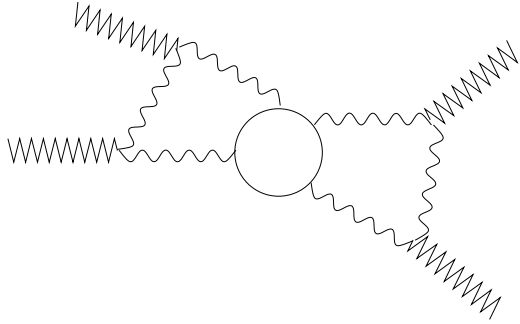,width=.4\linewidth}
\caption{A QED loop contribution to $\Gamma_{eff}[g_{\mu \nu}]$. Wavy lines 
are photons and solid lines are electrons.}
\end{figure} 

\begin{figure}[t]
\centering
\epsfig{file=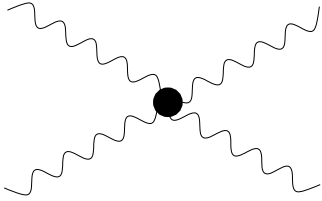,width=.25\linewidth}
\caption{Local effective light-by-light scattering vertex for soft photons.}
\end{figure}

Let us consider a concrete example from QED coupled to soft gravitons, Fig. 5.
We cannot throw away this whole contribution to the gravity effective action because 
even for soft graviton external lines there are imaginary parts which unitarize 
soft gravitons $\rightarrow$ soft photons processes (and their reverse), as well as 
soft photon-photon scattering due to the electron loop. On the other hand if we keep 
this diagram, we get a contribution to the cosmological constant set by the electron 
mass, which is  too big. The way too disentangle the hard and soft SM 
contributions is to simply do effective field theory below $\Lambda_{grav} \ll m_e$, 
where
 the soft photons ``see'' the electron loop as a local 
vertex, Fig. 6, the leading behaviour being of the rough form, 
\begin{equation}
{\cal L}_{eff} \ni \alpha_{em}^2 \frac{F^4}{m_e^4},
\end{equation}
$F_{\mu \nu}$ being the electromagnetic field strength.
Thus we recover all the same soft graviton and photon imaginary parts of Fig. 5
 with the diagram of Fig. 7, but now the contribution to the cosmological constant 
vanishes when we compute with dimensional regularization, since there is no mass 
scale in the propagators of the diagram itself, only in the overall coefficient.

\begin{figure}[t]
\centering
\epsfig{file=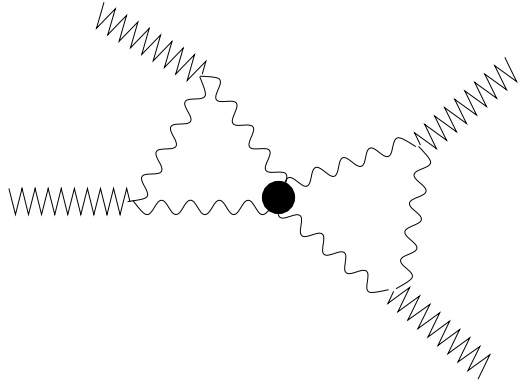,width=.4\linewidth}
\caption{An effective diagram with the same long distance physics as Fig. 5.}
\end{figure}

Thus a more precise statement is that with fat gravitons, massive SM 
($m_{SM} \gg \Lambda_{grav}$) and hard light SM pieces of  loop contributions to 
$\Gamma_{eff}[g_{\mu \nu}]$  may consistently be suppressed while soft light SM contributions are 
not. ``Consistent'' refers to the principles we have discussed before, unitarity and
GCI
relations in the regime we trust GR as an effective field theory now, 
namely  $< \Lambda_{grav}$, and locality down to $\ell_{grav}$. There are no 
robust contributions to the cosmological constant from mass scales above 
$\Lambda_{grav}$. 

\subsection{The Cosmological Constant in Fat Gravity}

Below $\Lambda_{grav}$, we have a standard effective field theory of 
GR coupled to SM light states. Here, {\it all}
 these states behave in a point-like manner. At the edge of this effective theory 
there are $\Lambda_{grav}$-mass vibrational excitations of the fat graviton, since in 
relativistic theory an extended object cannot be rigid. At least in 
this standard effective field theory regime 
 the quantum contributions to the cosmological constant should follow by the usual
 power-counting, that is $\sim {\cal O}(\Lambda_{grav}^4/16\pi^2)$. 
 The details however  depend on the details of the fat graviton. 
 Thus in a fat graviton theory naturalness 
implies that  the lower bound on the full cosmological constant is 
$\sim {\cal O}(\Lambda_{grav}^4/16\pi^2)$.

\section{Experimental/Observational Constraints/Predictions}

When we apply the above naturalness bound for a fat graviton to the 
observed dark energy \cite{ccexpt} \cite{pdg}, we can derive a bound on the graviton size, 
\begin{equation}
\ell_{grav} > 20 ~{\rm microns}.
\end{equation}
Of course since the naturalness bound involved an order of magnitude estimate for the
fat graviton quantum corrections to the cosmological constant, the bound on 
$\ell_{grav}$ is not  an exact prediction. However, it is a reasonably sharp 
prediction because 
much of the uncertainty is suppressed upon taking the requisite fourth root of the 
dark energy. If we can experimentally exclude $\ell_{grav}$ being in this regime we 
can falsify the idea that the fat graviton is (part of) the solution to the 
CCP.

\begin{figure}[t]
\centering
\epsfig{file=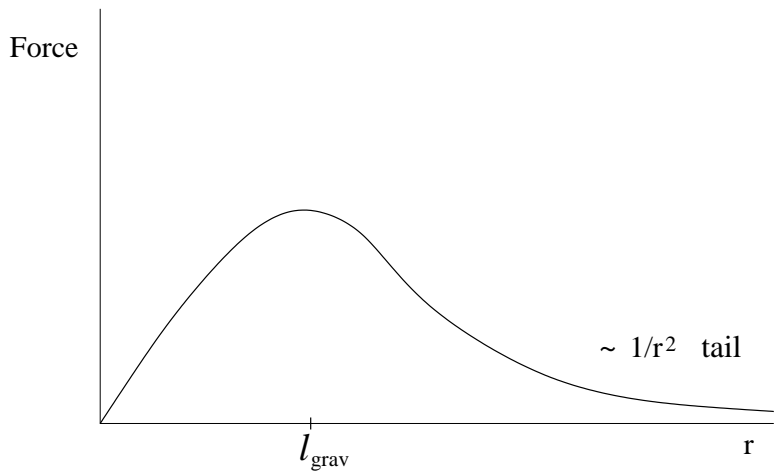,width=.6\linewidth}
\caption{The qualitative short-distance modification of Newton's Law due to a 
fat graviton of size $\ell_{grav}$.}
\end{figure}

How can we probe $\ell_{grav}$? This is the finite range of the vibrational excitations
of the fat graviton. They can therefore mediate deviations from Newton's Law at 
or below $\ell_{grav}$. The precise details of the deviations are sensitive to the 
unknown fat graviton theory, but the qualitative behavior follows from the essential
feature of 
the fat graviton as unable to mediate momentum transfers harder than $\Lambda_{grav}$.
In position space this implies Fig. 8, with the 
gravitational force being suppressed at distances below $\ell_{grav}$.
Of course this would be a striking signal to observe in sub-millimeter tests of gravity!
It sharply 
contrasts with the rapid short distance enhancement expected in theories with (only) 
large extra dimensions for gravity \cite{add}. Indeed there is no natural way (in the absence 
of supersymmetry \cite{scherk}) 
to have such suppression within standard point-like effective field theory.
Present bounds from such tests \cite{gravexpt1}   demonstrate that 
$\ell_{grav} < 0.2$ mm, so there is a fairly narrow regime to be explored in order to 
rule in or rule out a fat graviton approach to the CCP.

\section{Soft Graviton Effective Theory (SGET)}

\subsection{Basic Notions}

Can the graviton really be fat when SM matter is pointlike? Of course the only 
way to confidently answer in the affirmative is to build a consistent fundamental 
theory of this type. It is presently not known how to do this, say within 
string theory, the only known fundamental theory of any type of quantum gravity, 
where $\ell_{grav} = \ell_{string}$. One might worry that there is some sort of 
theorem anwering in the negative, that in a regime of pointlike matter the graviton 
must  be pointlike too. Such a theorem would have to show that our macroscopic 
tests of GR and microscopic tests of SM quantum field theory 
imply robust (pointlike) features of microscopic gravity. The best way to argue 
against such a theorem is to construct a consistent 
 effective field theory description which 
encapsulates precisely the present asymmetric 
experimental regimes for gravity and matter, but which does not commit itself to 
the details of microscopic gravity such as whether the graviton is pointlike. 
In particular, it would not suffer the usual CCP. 
We now turn to such a construction, generalizing  the methods of heavy 
particle effective theory \cite{hpet}. Our discussion is closely related 
 to the analysis of soft graviton couplings in Ref. \cite{wsoft}.
Earlier discussions of 
heavy 
particle effective theory applied to GR appear in Refs. \cite{donohue} and \cite{97}.

SGET is constructed from the fat graviton's ``point of view''. 
While the fat graviton can itself only mediate soft momentum transfers, it can 
``witness'' and couple to hard momentum transfers taking place within the SM sector. 
The momentum of a  freely propagating SM particle  can be expressed in terms of its 
$4$-velocity, 
\begin{equation}
p_{\mu} = m v_{\mu}, ~ v^2 = 1, ~ v_0 \geq 1. 
\end{equation}
Interactions with fat gravitons can change this momentum 
but by less than $\Lambda_{grav}$, 
\begin{equation}
\label{momsplit}
p_{\mu} = m v_{\mu} + k_{\mu}, ~  |k_{\mu}| < \Lambda_{grav}. 
\end{equation}
The basic idea of the effective  theory is to integrate out all 
 components of the SM field which are not of this form, that is far 
off-shell. The result is all that the  fat graviton can ``see'' and couple to. 
In this
regime, many SM loop effects which used to appear non-local in spacetime will appear 
local to the fat graviton. When coupling to gravity we will consider integrated out
 all massive vibrational
excitations of the fat graviton, leaving only the soft massless graviton mode. 
We know that GCI is a necessary feature for 
basic consistency of the couplings  of 
gravitons softer than $\Lambda_{grav}$ \cite{protect}. The IR use of GCI will be enough to recapture 
 the standard macroscopic tests of GR. Integrating out only SM 
fields which are far off-shell will not eliminate hard but on-shell SM particles, 
and therefore when properly matched, the effective theory should reproduce the 
high energy S-matrix of the SM as well. That is, the effective theory must
reproduce what we have seen of Nature, hard matter coupled to soft gravity. 

A very important distinction should be made. One might consider for example a proton,
propagating along, only interacting with gravity. In the effective field theory, we must
treat the proton as an elementary particle, not a composite of far off-shell quarks and 
gluons. All these off-shell particles are integrated out but the elementary proton is 
integrated ``in'' to match the usual SM physics. Intuitively, the fat graviton cannot 
distinguish the substructure of the proton. In this manner, there will be many more 
``elementary'' fields in the effective field theory than in the usual SM 
quantum field theory. 

In this paper a simplified problem is considered, where the SM is replaced by a toy model 
consisting of a single massive real scalar field, $\phi$, with $\lambda 
\phi^4$ 
coupling. 
The scalar has its own ``hierarchy'' fine-tuning problem, but we will ignore this 
irrelevant issue here. The central limitation is not the restriction to spin 0 (non-zero 
spin is tedious but straightforward), but rather the absence of light SM particles. We 
will rectify this omission elsewhere. We will refer to the renormalizable toy model 
as the ``SM'' or ``fundamental theory'' for the remainder of this section,
hopefully without causing confusion. We will also defer here 
the consideration of soft graviton quantum corrections, where soft graviton lines carry 
loop momenta. Again this will be presented elsewhere.

\subsection{Single Heavy Particle Effective Theory}

To begin, let us consider a state consisting  of a  single massive $\phi$-particle. 
We describe it with an effective field 
$\phi_v(x)$ which respects the split of momentum in Eq. (\ref{momsplit}). The 
fact that the $4$-velocity $v$ is formally unaffected by gravity means that it is 
just an unchanging label for the field, while the Fourier components of $\phi_v(x)$
 correspond to the ``residual'' momentum, $k_{\mu} < \Lambda_{grav}$,
 that alone can fluctuate with gravity interactions. However the split of Eq. (\ref{momsplit})
has an inherent redundancy formalized in terms of a  symmetry 
known as reparametrization invariance (RPI) \cite{rpi}: 
\begin{eqnarray}
v &\rightarrow& v + \delta v \nonumber \\
k &\rightarrow& k - m \delta v, 
\end{eqnarray}
where $\delta v$ is an infinitesimal change in velocity, $\delta v.v = 0$.
Obviously this transformation results in the same physical momentum $p$ and 
therefore the effective theory must identify the pairs $(v,k)$ and 
$(v + \delta v, k - m \delta v)$. 

Let us begin in flat space, without gravity. 
In terms of the effective field, RPI requires the identification 
\begin{equation}
\label{rpieq}
\phi_v(x) \longleftrightarrow e^{i m \delta v.x} \phi_{v + \delta v}(x).
\end{equation}
The simplest way to implement this is to treat this RPI as a
 ``gauge'' symmetry and ensure that the effective lagrangian is 
invariant under it. One must then be careful to only choose one element of 
any ``gauge orbit'' when extracting physics. 
It is straightforward to see that a covariant derivative given by 
\begin{equation}
D_{\mu} = \partial_{\mu} + i m v_{\mu}
\end{equation}
is required in order to build RPI effective lagrangians. For an isolated SM particle we 
need only consider quadratic lagrangians. Using the fewest derivatives (corresponding 
to the fewest powers of $k_{\mu}/m < \Lambda_{grav}/m$) we have 
\begin{eqnarray}
\label{freeeff}
{\cal L}_{eff} &=& \frac{1}{2m} |D_{\mu} \phi_v|^2 - \frac{m}{2} \phi^{\dagger}_v 
\phi_v \nonumber \\
&=& \phi^{\dagger}_v [iv.\partial + \frac{\partial^2}{2m}] \phi_v, 
\end{eqnarray}
where we integrated by parts to get the last line. In the first line we chose  a 
particular linear combination of two RPI terms. The overall coefficient is just a 
conventional wavefunction renormalization, but the relative coefficient is chosen to 
satisfy the physical requirement that the propagator have a pole at $k = 0$, given 
the interpretation of Eq. (\ref{momsplit}). 

Formally, in the 
derivative expansion ($k/m$ expansion) the dominant term in ${\cal L}_{eff}$ is 
$ \phi^{\dagger}_v iv.\partial \phi_v$, while the subleading term, 
$\phi^{\dagger}_v \frac{\partial^2}{2m} \phi_v$ can be treated perturbatively, 
that is, as a higher derivative ``interaction'' vertex. The effective propagator 
is then given by 
\begin{equation}
\label{prop}
\frac{i}{k.v + i \epsilon}.
\end{equation}
 This treatment will suffice for most of the 
examples given below. An important property 
of the effective propagator is that it contains a single pole, rather than the two
poles at positive and negative energy of a standard field theory 
propagator.  The reason
is simple to understand. Given that (when gravity is finally included) the maximum 
residual momentum $k$ is $< \Lambda_{grav}$, even though the sign of $k_0$ is not fixed
the sign of the  total energy $p_0$ is clearly positive. Fat gravitons cannot 
impart the momentum transfers needed to go near the usual negative energy pole. 
Without negative total energies the effective field is necessarily complex, as 
indicated, and $\phi_v$ is purely a creation operator while $\phi_v^{\dagger}$ is 
purely a destruction operator. Such a split would look non-local in a fundamental 
quantum field theory, but not in the SGET ``seen'' by  the fat graviton. 

Sometimes one studies processes where components 
of $k$ orthogonal to $v$ are larger than $k.v$, such that one cannot treat 
$\phi^{\dagger}_v \frac{\partial^2}{2m} \phi_v$ perturbatively. It would seem then 
that if one uses all of Eq. (\ref{freeeff}) to determine the propagator one would 
find two poles again. However, in these circumstances one only needs to resum the 
$\partial_{\perp v}^2/2m$ part of the $\partial^2/2m$ term in the propagator. The 
other $(v.\partial)^2/2m$ piece of $\partial^2/2m$ would then be of even higher
order and could still be treated as a perturbtation. In this way the 
 resulting propagator, 
\begin{equation}
\label{nlprop}
\frac{i}{k.v + k_{\perp v}^2/2m + i \epsilon},
\end{equation}
again has a single pole. We will see such an example in what follows. 

Let us now couple soft gravity to ${\cal L}_{eff}$. Interactions for soft 
massless spin-2 modes 
of the fat graviton only make sense if protected by GCI, so our 
effective theory must be exactly GCI. 
The only assumption is that there is a sensible theory of 
a fat object with a massless spin-2 mode coupling to matter. 
To determine the possible couplings
we must first determine the spacetime transformation properties of $\phi_v(x)$. 
Naively, one would guess that since the particle has spin 0, that in flat 
space $\phi_v(x)$ is a scalar field of Poincare invariance, and becomes a GCI scalar
field once we turn on gravity. However the first presumption is incorrect.
To see this consider
a flat space Poincare transformation defined by 
\begin{equation}
x^{\mu} \rightarrow \Lambda^{\mu}_{~ \nu} x^{\nu} + a^{\mu}.
\end{equation}
Restricting to an {\it infinitesimal} transformation of this type we have 
\begin{eqnarray}
\phi_v(x) &\rightarrow& \phi_{\Lambda v}(\Lambda x + a) \nonumber \\
&=& \phi_{v + \delta v}(\Lambda x + a) \nonumber \\
&\equiv& e^{- i m \delta v.(\Lambda x + a)} \phi_{v}(\Lambda x + a),
\end{eqnarray}
where in the second line we have used the fact that for an infinitesimal Lorentz 
transformation we can always write $\Lambda v$ in the form $v + \delta v$ where 
$v.\delta v = 0$, and in the last line we have used the RPI equivalence relation, 
Eq. (\ref{rpieq}). This is not the transformation property of a Poincare scalar field.
But clearly $e^{i m v.x} \phi_v(x)$ {\it is} a 
Poincare scalar. When we couple to gravity, it is this combination that remains a 
scalar field. 

A generally covariant derivative of the scalar is easy to form, 
\begin{equation}
\partial_{\mu} [e^{i m v.x} \phi_v(x)] = e^{i m v.x} (\partial_{\mu} + i m v_{\mu}) 
\phi_v(x). 
\end{equation}
Thus the combination $D_{\mu} \equiv 
(\partial_{\mu} + i m v_{\mu})$ here is forced on us by both GCI 
and RPI. The leading GCI and RPI effective lagrangian is then given by 
\begin{eqnarray}
{\cal L}_{eff} &=& \sqrt{-g} \{ 
\frac{g^{\mu \nu}}{2m} D_{\mu} \phi_v^{\dagger} 
D_{\nu} \phi_v - \frac{m}{2} 
\phi^{\dagger}_v \phi_v \}.
\end{eqnarray}
To leading order in the expansions in $k/m$ and $m/M_{Pl}$, this yields
\begin{equation}
{\cal L}_{eff} = \phi^{\dagger}_v \{ i v.\partial - \frac{m}{2 M_{Pl}} 
v_{\mu} v_{\nu} h^{\mu \nu} \} \phi_v,
\end{equation}
which reproduces the standard equivalence of gravitational and inertial mass. 

Note that in our derivation of this equivalence we did not make use of the existence 
of standard GR at short distances, even though 
short distance physics may well contribute in complicated ways 
to the mass of the SM state.

Let us now ask what robust loop contribution the effective theory makes to the 
cosmological constant. Since the effective theory does have GCI we can just 
calculate the pure SM vacuum energy with no graviton external legs. Apparently this 
requires us to interpret the expression,
\begin{equation}
\int d^4 k ~ \ln (k.v + i \epsilon).
\end{equation}
Note that there is no physical SM mass scale in this expression so there is no 
robust contribution from known physics here at all even though the effective theory 
does reproduce the coupling of  massive SM particles to gravity. We can simply set the 
above expression to zero. This can be thought of as normal ordering. The reason for 
having no robust contribution to the cosmological constant is because the effective 
theory of the soft gravitons does not know whether the graviton is fat or not, or 
indeed whether the heavy matter is highly composite, say the Earth (if the gravitons 
are very soft), or solitonic like a 0-brane in string theory. 
All these possibilites lead to and effective theory of the same form. The effective theory
 cannot commit to (even the rough size of)
a cosmological constant contribution without knowing the differentiating physics which lies 
beyond itself.

\subsection{Effective theory with SM Interactions}

Let us now consider how to generalize the effective theory for processes involving
 several interacting SM particles coupled to soft gravitons. 
The  SGET general form can be compactly expressed,
\begin{equation}
\label{generaleff}
{\cal L}_{eff} = \sqrt{-g} ~ \sum_{v, v'} 
\frac{\kappa_{v v'}(D_{\mu}/m)}{m^{3/2(N + N') - 
4} } ~
\phi^{\dagger}_{v'_1} ... \phi^{\dagger}_{v'_{N'}} ~ \phi_{v_1} ... \phi_{v_N}
e^{i m (\sum v - \sum v').x}, 
\end{equation}
where there are dimensionless coefficients 
$\kappa_{v v'}$ which in general can contain 
GCI  and RPI derivatives acting on any of the effective fields, contracted with the 
inverse metric, $g^{\mu \nu}$. Of course the non-trivial parts of 
such derivatives correspond to residual momenta, $k$, balanced by powers of $1/m$. 
Therefore they are relevant for subleading effects in $\Lambda_{grav}/m$. 

The general procedure for specifying the (leading terms) of the SGET is to match 
the effective theory to the fundamental SM in the absence of gravity, and then to 
covariantize minimally with respect to (soft) gravity. The matched correlators are 
those with nearly on-shell external lines, Eq. (\ref{momsplit}).
This will reproduce the soft 
graviton amplitudes of the fundamental SM directly coupled to gravity in the 
standard way, but now without any reference to pointlike graviton couplings. Therefore 
it is compatible with a having a fat graviton whose massless mode is protected only 
by an infrared GCI. Note that the hard SM momentum transfers are to be 
described in the effective theory by $v_i \rightarrow v'_j$, that is by a change of labels. 
Changes in effective field momenta, that is residual momenta, are necessarily soft. 
This is how a fat graviton sees SM processes, the hard momentum transfers are a given 
feature of such processes which the fat graviton cannot influence. See Ref. \cite{iramike} for an
effective field theory formalism exhibiting similar dynamical label-changing 
in the context of non-relativistic QCD.

It is unusual to see explicit $x$-dependence in (effective) lagrangians such as appears 
in the phase factor. Here it is  required by overall momentum conservation, rather than 
conservation of residual momenta. More formally, it is required by RPI, as well as GCI 
(recalling that it is $\phi_v e^{i m v.x}$ which is a  scalar of GCI). It is 
possible, and perhaps prettier, to partially gauge fix the RPI, by reparametrizing
 (as can always be done) such that  $\sum v = \sum v'$, so that the 
phase factor $\rightarrow 1$ without compromising GCI. We will call this the 
``label-conserving gauge''.
However, when loops are considered
we will find it 
 convenient (but not necessary) to depart infinitesimally from this gauge fixing and 
consider an infinitesimal phase. 

Below, we will work to leading order about the limits $M_{Pl}, m, \Lambda_{grav} \rightarrow 
\infty, ~ \Lambda_{grav}/m \ll 1,$ with $m/M_{Pl}$ fixed. 
This formal limit simplifies the effective theory. It is similar to doing 
standard effective 
field theory calculations, including matching, without an explicit 
cutoff, even though physically one imagines new physics cutting off the effective theory
at a finite scale. 

\subsubsection{Tree-level Matching}

\begin{figure}[t]
\centering
\epsfig{file=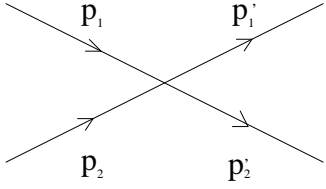,width=.25\linewidth}
\caption{Two-particle tree-level scattering. Arrows indicate incoming and 
outgoing nearly on-shell external states.}
\end{figure} 

Let us follow the general procedure outlined above and first shut off gravity and work
in flat space.
Consider the tree level diagram for $2 \rightarrow 2$ scattering 
in the fundamental theory, Fig. 9, where the external momenta are nearly on-shell.
We can then express these external momenta in the form of Eq. (\ref{momsplit}), where
we choose label-conserving gauge, 
\begin{equation}
v_1 + v_2 = v'_1 + v'_2.
\end{equation}
This amplitude is then straightforwardly matched by including a $2 \rightarrow 2$ 
effective vertex, 
\begin{equation}
\label{eff4}
{\cal L}_{eff} \ni \frac{\lambda}{4m^2} \phi^{\dagger}_{v_1'} \phi^{\dagger}_{v_2'} ~
\phi_{v_1} \phi_{v_2}. 
\end{equation}
The $1/4m^2$ factor only arises due to the different normalizations of the interpolating 
fields between the fundamental and effective theories.

\begin{figure}[t]
\centering
\epsfig{file=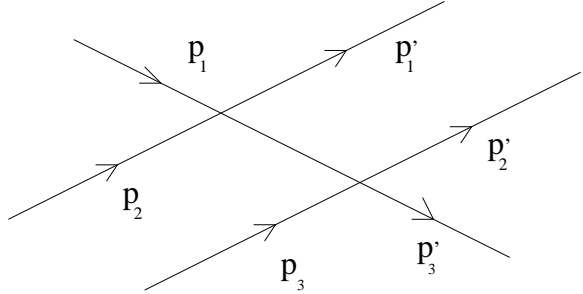,width=.45\linewidth}
\caption{A three-particle scattering diagram in the fundamental theory.}
\end{figure}

\begin{figure}[t]
\centering
\epsfig{file=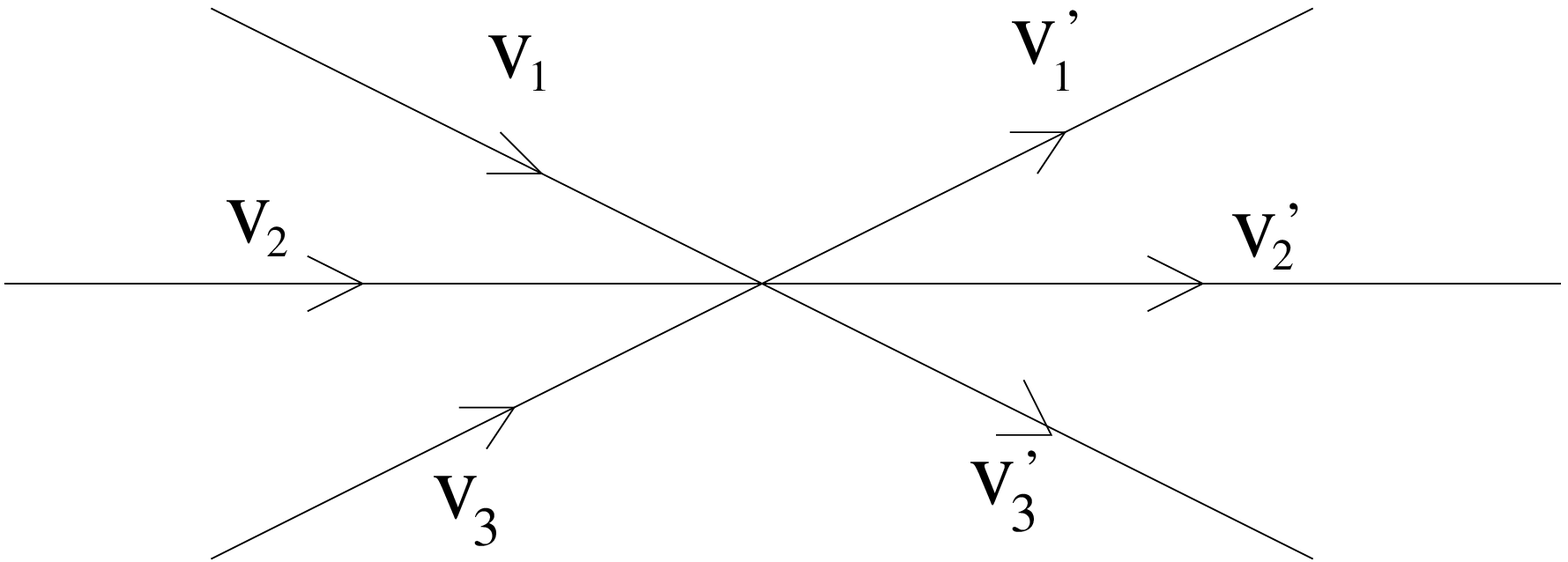,width=.45\linewidth}
\caption{Effective vertex obtained by integrating out the far off-shell 
internal line in Fig. 10.}
\end{figure}

Next, consider the $3 \rightarrow 3$ process in the fundamental theory, Fig. 10. 
Again, the external lines are nearly on-shell, so we can express them as 
\begin{equation}
p_i = m v_i + k_i, ~ p_i' = m v'_i + k'_i, ~  i = 1,2,3, 
\end{equation}
with label conservation. The internal line has momentum, 
\begin{equation}
p_{int} = m(v_1 + v_2 - v_1') + k_1 + k_2 - k_1'.
\end{equation}
Generically in such hard SM collisions, in the limit $\Lambda_{grav}/m \ll 1$, the 
internal lines will be far off-shell and to leading order we can drop the $k$'s,
\begin{equation}
p_{int} \approx m(v_1 + v_2 - v_1').
\end{equation}
We can then match the fundamental diagram with an effective vertex, Fig. 11, given by 
\begin{equation}
\label{eff6} 
{\cal L}_{eff} \ni \frac{\lambda^2}{8m^5[(v_1 + v_2 - v_1')^2 - 1]} ~ 
\phi^{\dagger}_{v_1'} \phi^{\dagger}_{v_2'} \phi^{\dagger}_{v_3'} ~
\phi_{v_1} \phi_{v_2} \phi_{v_3}. 
\end{equation}

Notice that what was a fundamentally non-local exchange requiring an off-shell 
internal line in the fundamental theory, that is with non-analytic dependence on 
the external total momenta, is replaced in the effective vertex by a local interaction
with a non-analytic dependence only on the labels, $v, v'$. Physically, this is because
the process is fundamentally non-local, but is local down to $\Lambda_{grav} \ll m$, 
that is local ``enough'' for a fat graviton. 

If we had worked to higher order in $\Lambda_{grav}/m$, matching would have 
resulted in higher derivative effective vertices, corresponding to having retained 
higher powers of $k/m$ in expanding $1/(p_{int}^2 - m^2)$ for small $k < \Lambda_{grav}$.

For processes of the form of Fig. 10, there are also exceptional situations which 
result in $p_{int}$ being nearly on-shell. These arise when one considers experiments 
(in position space) where the interaction region for wavepackets of particles $1$ and 
$2$ is greatly displaced from the trajectory of the wavepacket for particle $3$, 
compared with the size of the wavepackets. Thus the three-particle scattering 
is dominated by a sequence of two-particle scatterings, 
\begin{eqnarray}
p_1 + p_2 &\rightarrow& p'_1 + p_{int} \nonumber \\
p_3 +  p_{int} &\rightarrow& p'_2 + p'_3, 
\end{eqnarray}
where all momenta are nearly on-shell. In these exceptional cases we can 
express $p_{int} = m v_{int} + k_{int}$ with label conservation:
\begin{eqnarray}
v_1 + v_2 &=& v'_1 + v_{int} \nonumber \\
v_3 +  v_{int} &=& v'_2 + v'_3.  
\end{eqnarray}
In the limits we are considering, the fundamental internal propagator then approaches 
the effective propagator, Eq. (\ref{prop}), 
up to the convention-dependent field normalization, 
\begin{eqnarray}
\frac{1}{ p_{int}^2 - m^2 + i \epsilon} &=& \frac{1}{2m} ~ \frac{1} {k_{int}.v_{int} + 
k_{int}^2/2m + i \epsilon} \nonumber \\
&\approx& \frac{1}{2m} ~ \frac{1} {k_{int}.v_{int} + i \epsilon}.
\end{eqnarray}
Thus the exceptional fundamental diagram of the form of Fig. 10 is matched 
by an effective diagram of the same form, but where we use the effective $4$-point 
vertices matched earlier, 
\begin{equation}
{\cal L}_{eff} \ni \frac{\lambda}{4m^2} \phi^{\dagger}_{v_1'} \phi^{\dagger}_{v_{int}} ~
\phi_{v_1} \phi_{v_2} + \frac{\lambda}{4m^2} \phi^{\dagger}_{v_2'} \phi^{\dagger}_{v_3'} ~
\phi_{v_3} \phi_{v_{int}}, 
\end{equation}
and the effective propagator, Eq. (\ref{prop}), for the internal line.

More general tree level amplitudes are generally matched by a
combination of the two procedures illustrated above, 
introducing new effective vertices and connecting effective vertices with effective 
propagators.
Tree-level unitarity in the effective theory arises from the imaginary parts of 
amplitudes due to the $i \epsilon$-prescription when internal lines go on shell, precisely
 matching the fundamental theory. 

There is a situation one can imagine for finite $\Lambda_{grav}$ where we carefully tune 
the external momenta on Fig. 10 so that the internal line is intermediate between the 
two (more generic) situations we have considered of being far off-shell or nearly on-shell, 
that is, the internal line is of order $\Lambda_{grav}$ off-shell. 
In that case our simple procedure does not always 
allow the fundamental graph to be matched by a 
local effective vertex or an effective theory graph. It appears that there is a more 
complicated scheme for matching even these cases within SGET, and that they pose a technical  
rather than conceptual challenge. We will study these cases elsewhere.

\subsubsection{Coupling the effective matter theory to soft gravity}

Coupling the vertices of ${\cal L}_{eff}$ to gravity is very simple. For example, 
the minimal covariantization of  Eq. (\ref{eff6}) is given by multiplying by 
$\sqrt{-g}$. If we had worked to higher order in the derivative expansion we would have 
to also covariantize these derivatives with respect to GCI and contract them using 
$g^{\mu \nu}$.

\begin{figure}[t]
\centering
\epsfig{file=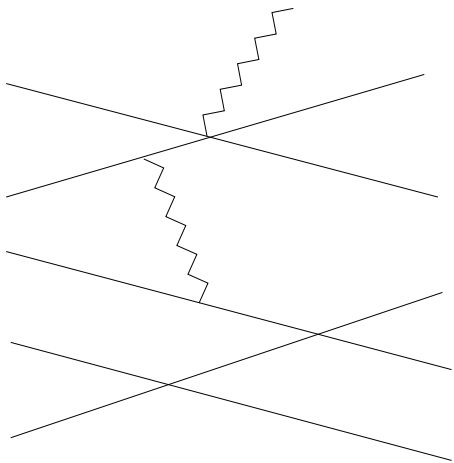,width=.35\linewidth}
\caption{A soft graviton $+$ hard SM diagram in the effective theory.}
\end{figure} 

We can now use the GCI effective theory to compute hard SM processes coupled to soft 
gravitons, such as say Fig. 12. The results automatically match with the leading 
behavior of the analogous 
diagrams if we coupled soft gravitons directly to  the fundamental theory in the usual 
way. There is no extra tuning of couplings needed to recover standard gravitational 
results beyond imposing infrared GCI. The dominant couplings of soft gravitons therefore
do not distinguish  whether (a) the graviton is fat and GCI is only a guiding symmetry of 
the couplings in the far infrared, or (b) the gravitons are point-like and GCI  
governs their couplings to the fundamental theory in the standard way usually 
assumed.


\subsubsection{Matching Loops}

\begin{figure}[t]
\centering
\epsfig{file=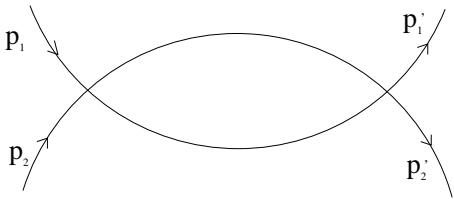,width=.35\linewidth}
\caption{A simple loop diagram of the fundamental SM 
theory with nearly on-shell 
external lines.}
\end{figure} 

\begin{figure}[t]
\centering
\epsfig{file=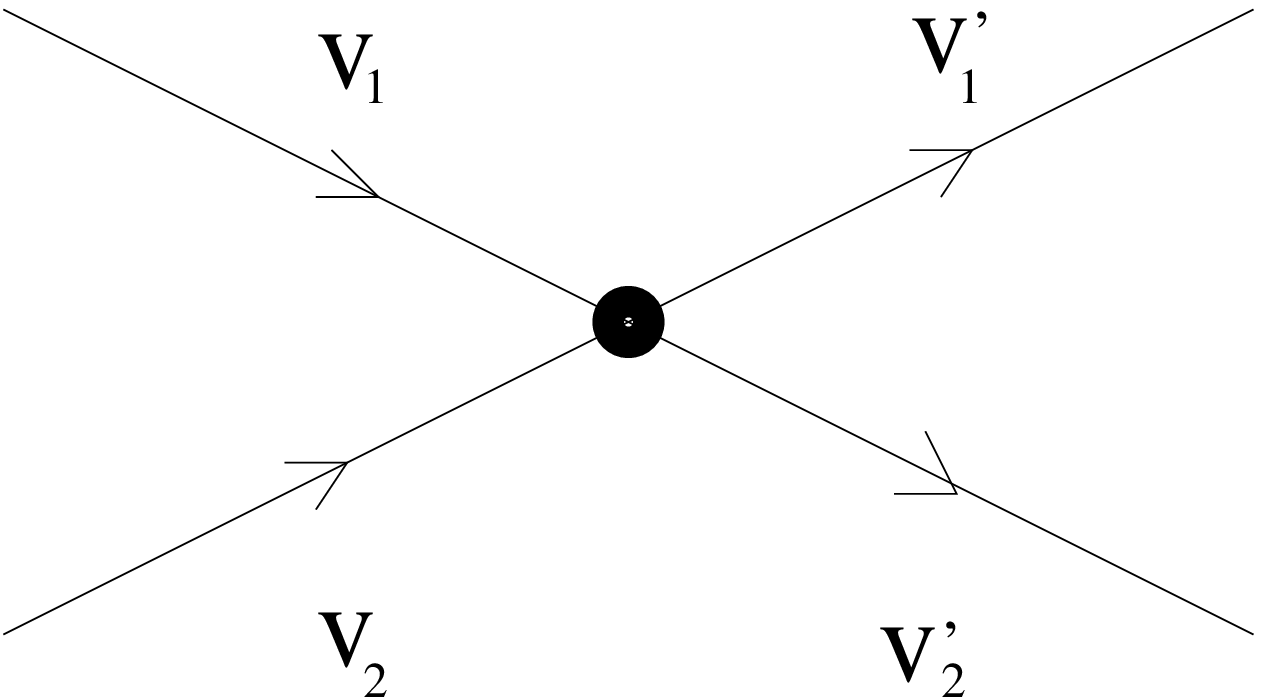,width=.25\linewidth}
\caption{The form of a one-loop correction to the four-point effective vertex.
}
\end{figure} 

Here, we will match the simplest fundamental non-trivial loop diagram, Fig. 13, in flat space. 
It 
illustrates the essential new complication that loops bring in the presence of a 
``gauge'' symmetry like RPI, namely the need to gauge fix and determine the right 
integration measure. We will first consider the case where the incoming momenta are 
far away from the two-particle threshold. We can decompose the momenta as usual, 
\begin{equation}
p_i = m v_i + k_i, ~ p_i' = m v'_i + k'_i, 
\end{equation}
with label conservation. Denote the loop amplitude by $\Gamma_{fund}(p, p')$.
There is a general result for Feynman diagrams \cite{bj}, which is 
straightforward to explicitly check in this example, that $\Gamma_{fund}$ is 
locally analytic in the external momenta except when near a threshold, which we 
assumed above is not the case. That is, $\Gamma_{fund}(mv + k, mv' + k')$ is analytic in 
$k$ and $k'$ and has a series expansion. Naively, we might try to match the whole
diagram in the effective theory by Fourier transforming this series expansion into 
local operators in the effective theory. However, we cannot do this as it violates 
hermiticity of the effective lagrangian and ultimately unitarity. To see this focus on 
the leading term in the expansion and how it would appear as an effective vertex, 
depicted in Fig. 14, 
\begin{equation}
\label{naivevx}
{\cal L}_{eff} \ni \sum_{v, v'} \frac{\Gamma_{fund}(v, v')}{4m^2} \phi^{\dagger}_{v_1'} 
\phi^{\dagger}_{v_2'} ~ \phi_{v_1} \phi_{v_2} ~ ? 
\end{equation}
Hermiticity of the effective lagrangian requires that the coupling $\Gamma_{fund}(v, v')$ be 
real, but by unitarity in the fundamental theory or direct calculation we know that 
$\Gamma_{fund}(v, v')$ has an imaginary part corresponding to the region of integration 
where the internal propagators are on-shell. Of course we are free to replace 
$\Gamma_{fund}(v, v') 
\rightarrow {\rm Re} ~ \Gamma(v, v')$ in Eq. (\ref{naivevx}), but then 
the imaginary piece must be matched from another source. 

Obviously, in the 
 effective theory there is also a diagram of the form of Fig. 13, but where 
the vertices and propagators are replaced by  effective vertices, Eq. (\ref{eff4}), and 
propagators, Eq. (\ref{prop}). There are two internal momenta now, 
\begin{eqnarray}
p_{int} &=& mv_{int} + k_{int} \nonumber \\
p'_{int} &=& mv'_{int} + k'_{int},
\end{eqnarray}
so naively the loop momentum integration measure has the form
\begin{equation}
\sum_{v_{int}} \sum_{v'_{int}} \int d^4 k_{int} \int d^4 k'_{int} ~
\delta^4(mv_{int} + k_{int} + mv'_{int} + k'_{int} - p_1 - p_2).
\end{equation}
This is ill-defined, there are too many sums going on because we are  
multiple-counting combinations $(v,k)$ that should be indentified by RPI. 
That is the correct measure has the form 
\begin{equation}
\frac{\sum_{v_{int}} \sum_{v'_{int}} \int d^4 k_{int} \int d^4 k'_{int}}{RPI} ~ 
\delta^4(mv_{int} + k_{int} + mv'_{int} + k'_{int} - p_1 - p_2),
\end{equation}
where the denominator means to identify RPI related combinations. Our job is to do
this by gauge fixing RPI, so that we are summing just one representative of each RPI 
equivalence class. This is the central subtlety in computing with the effective 
theory at loop level. 

We will fix the following gauge. The total incoming momentum $p_1 + p_2$ is necessarily 
timelike. We will define our coordinates in its rest frame for convenience, not because 
the gauge fixing breaks manifest relativistic invariance (which in any case would not be a 
disaster if properly treated). In this frame we will gauge fix 
\begin{equation}
\vec{k}_{int} = \vec{k}'_{int} = 0,
\end{equation}
that is, only $k^0_{int}, k^{'0}_{int} \neq 0$. 
It is obvious that any nearly on-shell 
momenta like those in the internal lines of effective theory diagrams, 
$p_{int}, p'_{int}$, can be decomposed in this gauge, 
\begin{eqnarray}
\label{gfix}
\vec{v} &\equiv& \frac{\vec{p}}{m} \nonumber \\ 
v_0 &\equiv& \sqrt{1 + \vec{v}^2} \nonumber \\
k_0 &\equiv& p_0 - mv_0.
\end{eqnarray}
Thus, an internal line is now specified by four real numbers, $k_0, \vec{v}$ rather 
than seven, $k_{\mu}, v_{\mu}: v^2 = 1$. This is the right counting. We can get the 
correct RPI measure of integration for the internal momenta, by noting that obviously
$\int d^4 p_{int}$ is a RPI measure, since the $(v,k)$ split has not been made.
Using Eq. (\ref{gfix}) then leads to the RPI measure 
\begin{equation}
m^3 \int d^3 \vec{v}_{int} \int d k^0_{int}.
\end{equation}

Thus the effective theory version of Fig. 13 is given by,
\begin{eqnarray}
\Gamma_{SGET loop} &=& \frac{i \lambda^2}{(2 \pi)^4} ~  
\frac{\sum_{v_{int}} \sum_{v'_{int}} \int d^4 k_{int} 
\int d^4 k'_{int}}{RPI} ~ 
\delta^4(mv_{int} + k_{int} + mv'_{int} + k'_{int} - p_1 - p_2) \nonumber \\ 
&~& ~ \times \frac{1}{k_{int}.v_{int} 
+ i \epsilon} ~ \frac{1}{k'_{int}.v'_{int} + i \epsilon} \nonumber \\
&=& \frac{i \lambda^2 m^6}{(2 \pi)^4} ~ \int d^3 \vec{v}_{int} \int d k^0_{int} 
\int d^3 \vec{v}'_{int} \int d k'^{0}_{int} ~ 
\delta^3(m \vec{v}_{int}  + m \vec{v}'_{int} - \vec{p_1} - \vec{p_2}) \nonumber \\
&~& ~ \times
\delta(m v^0_{int} + m v'^{0}_{int} + k^0_{int} + k'^{0}_{int}
 - p^0_{1} - p^0_{2}) ~ 
\frac{1}{k^0_{int}v^0_{int}
+ i \epsilon} ~ \frac{1}{k'^{0}_{int}.v'^{0}_{int} + i \epsilon}.
\end{eqnarray}
Note that the two tree effective vertices here do not satisfy label conservation. 
This is because with our present gauge fixing it would be inconsistent to also 
insist on label-conserving gauge. Therefore we relax the latter requirement, which 
in any case has only a cosmetic value.

We will work to leading (zeroth) order in the {\it external} residual momenta, so that 
we can simply take $p_i = m v_i, p'_i = m v'_i$. In our choice of frame we then have
\begin{eqnarray}
\vec{v}_1 + \vec{v}_2 &=& 0 \nonumber \\
E_{tot}/2 &\equiv& v^0_{1} = v^0_{2}.
\end{eqnarray}
Substituting this in and integrating the $\delta$-functions gives,
\begin{eqnarray}
\Gamma_{SGET loop} &=& 
\frac{i \lambda^2 m^3}{(2 \pi)^4} \int \frac{d^3 \vec{v}_{int}}{(v^0_{int})^2} 
\int d k^0_{int} ~  
\frac{1}{k^0_{int}
+ i \epsilon} ~ \frac{1}{E_{tot} - 2 m v^0_{int}  - k^0_{int} + i \epsilon}.
\end{eqnarray}
The $k^0_{int}$-integral is finite and done by contour integration,, 
\begin{equation}
\Gamma_{SGET loop} = 
\frac{\lambda^2 m^2}{2 (2 \pi)^3} \int \frac{d^3 \vec{v}_{int}}{(v^0_{int})^2} 
\frac{1}{v^0_{int} - E_{tot}/2m  + i \epsilon}.
\end{equation}

Now, the remaining $\vec{v}_{int}$-integral representation of $\Gamma_{SGET loop}$ 
is logarithmically divergent just as the familiar $\Gamma_{fund}$. However, it is 
straightforward to see that the imaginary parts, related by unitarity to 
tree-level two-particle scattering, are finite and agree (up to the usual difference in  
normalization of states), 
\begin{equation}
{\rm Im} ~ \Gamma_{fund} = {\rm Im} ~ \Gamma_{SGET loop}/4m^2 = 
\frac{\lambda^2}{16 \pi^2} \int \frac{d^3 \vec{v_{int}}}{(2 v^0_{int})^2} 
\delta(v^0_{int} - E_{tot}/2m).
\end{equation}
This of course just corresponds to integrating over the phase space for on-shell 
$2$-particle intermediate states.
Thus the imaginary parts are matched between the effective and fundamental theories. 

It is the real parts which diverge. 
In both cases the integrals converge in a 
$(4 - \delta)$-dimensional spacetime, that is with dimensional regularization. The important 
point is that not just Re~$\Gamma_{fund}$, discussed above,
 but also Re $\Gamma_{SGET loop}$ are 
local analytic functions of the external momentum. The latter is easily seen by 
deformation of the integration contour for $|\vec{v}_{int}|$ to avoid the 
$E_{tot}/2m$ pole, as long as $E_{tot}/2m > 1$ as we assumed (that is we are above 
the two-particle threshold). Therefore $\Gamma_{SGET loop}$ is locally analytic in 
$E_{tot} = \sqrt{(m v_1 + k_1 + m v_2 + k_2)^2}$, that is, analytic in the $k_i$. 
Thus we can introduce a local {\rm and hermitian } effective vertex given by 
$4 m^2 {\rm Re} ~ \Gamma_{fund} - {\rm Re} ~\Gamma_{SGET loop}$, of the form of Fig. 14
to match the real parts, thereby completing the matching procedure.

Of course we can also reverse our starting assumption and reconsider matching if we 
chose $p_1 + p_2$ to be nearly at the two-particle threshold. 
In that case, we can 
decompose all the 
external particles of Fig. 13 to have a common 4-velocity label,
\begin{equation}
p_{i} = m v + k_i, ~ p'_{i} = m v + k'_i.
\end{equation}
The problem now is that even the real part of the diagram can be non-analytic in the 
$k_i$ and therefore cannot be captured by a local effective vertex, but must emerge 
from an effective diagram of the form of Fig. 13 just as the imaginary part must.
But in just this near-threshold case, this proves to be possible. This situation 
corresponds to a rather standard case in heavy particle effective theory. For example, 
see Ref. \cite{nucl}. Nevertheless, we will verify below that things work.

By a standard 
calculation we have 
\begin{eqnarray}
\Gamma_{fund} &\equiv& \frac{i \lambda^2}{(2 \pi)^4} \int d^4 p_{int}
~ \frac{1}{p_{int}^2 - m^2 + i \epsilon} ~ \frac{1}{(p_1 + p_2 - p_{int})^2 
- m^2 + i \epsilon} \nonumber \\
&=& \frac{\lambda^2}{16 \pi^2} \sqrt{ 1 - 4m^2/(p_1 + p_2)^2} ~ 
\ln \{ \frac{\sqrt{ 1 - 4m^2/(p_1 + p_2)^2} + 1}{\sqrt{ 1 - 4m^2/(p_1 + p_2)^2} - 1} \}
+ {\rm constant},
\end{eqnarray}
where the constant term contains the usual divergence. Matching the constant term is 
trivial so we will focus on the term non-analytic in the external momenta. 
We will work to leading non-trivial order in $k_i/m$, 
\begin{equation}
(p_1 + p_2)^2 = 4m^2 + 2mv.(k_1 + k_2) + (k_1 + k_2)^2 \approx 4m^2 + 2mv.(k_1 + k_2).
\end{equation}
Substituting into $\Gamma_{fund}$ and working to leading order yields
\begin{equation}
\Gamma_{fund} \ni \frac{i \lambda^2}{16 \pi} \sqrt{\frac{(k_1 + k_2).v}{m}}.
\end{equation}
Because we are near threshold this is not analytic in even the residual momenta. 
This allows it to have the behavior required by unitarity, imaginary above 
threshold, $(k_1 + k_2).v > 0$, but real below, $(k_1 + k_2).v < 0$.

We now compute the analogous loop diagram in SGET of 
the form of Fig. 13. Again we must gauge fix. The simplest procedure is to 
write
\begin{equation}
p_{int} = m v + k_{int}, ~ p'_{int} = m v + k'_{int},
\end{equation}
 with the {\it same} fixed velocity as the external lines. Thus the RPI integration 
measure is obvious, $\int d^4 p_{int} = \int d^4 k_{int}$. 
Let us first try to work strictly to leading order in the $1/m$ expansion. After 
integrating the momentum-conserving $\delta$-functions we have 
\begin{eqnarray}
\Gamma_{SGET loop} 
&=& \frac{i \lambda^2}{(2 \pi)^4} \int d^4 k_{int} ~ 
\frac{1}{k_{int}.v + i \epsilon} ~ \frac{1}{(k_1 + k_2 - k_{int}).v + 
i \epsilon} \nonumber \\
&=& \frac{\lambda^2}{(2 \pi)^3} ~ \frac{1}{(k_1 + k_2).v} ~ \int d^3 k_{int \perp v}
~ 1 \nonumber \\
&=& 0,
\end{eqnarray}
where in the second line we have done the finite $\int d (k_{int}.v)$ by contour 
integration. We see a cubic divergence. Formally, this corresponds to an ${\cal O}(
m/(k_1 + k_2).v)$ effect. However, in dimensional regularization 
$\int d^3 k_{int \perp v}
~ 1 = 0$. Therefore we must work to higher order in $1/m$ by 
keeping the dominant subleading terms in the propagators, as in Eq. (\ref{nlprop}),
\begin{eqnarray}
\Gamma_{SGET loop} 
&=& \frac{i \lambda^2}{(2 \pi)^4} \int d^4 k_{int} ~ 
\frac{1}{k_{int}.v + k_{int \perp v}^2/2m + i \epsilon} \nonumber \\
&~& ~ \times  
\frac{1}{(k_1 + k_2 - k_{int}).v + 
(k_1 + k_2 - k_{int})_{\perp v}^2/2m +  i \epsilon} \nonumber \\
&\approx& \frac{\lambda^2}{(2 \pi)^3} \int d^3 k_{int \perp v} ~  
\frac{1}{(k_1 + k_2 - k_{int}).v + 
k_{int \perp v}^2/m +  i \epsilon} \nonumber \\
&=& \frac{i m^2 \lambda^2}{4 \pi} \sqrt{\frac{(k_1 + k_2).v}{m}},
\end{eqnarray}
where in the second line we have again done the finite $\int d (k_{int}.v)$ by contour 
integration and kept only the leading terms in $(k_1 + k_2)/m$, and in the last line 
we have evaluated the linearly divergent $\int d^3 k_{int \perp v} $ using 
dimensional regularization. As can be seen, this precisely matches the non-analytic 
terms in $\Gamma_{fund}$ (taking into account the different state normalization as usual).


\subsection{The Cosmological Constant in SGET}

Let us finally consider what robust contributions to the infrared cosmological constant 
emerge within SGET. Because of the procedure for obtaining the SGET by {\it first} 
matching to the fundamental flat space theory {\it and then} covariantizing 
with respect to gravity, there is no cosmological constant term in the effective 
lagrangian,  Eq. (\ref{generaleff}). Of course we could simply add one, but there is no 
robust reason to do so except quantum naturalness. Therefore let us consider loop 
contributions to the infrared cosmological constant within SGET. We can make use 
of the GCI enjoyed by SGET to simply compute the pure vacuum diagrams in the complete 
absence of graviton lines, that is, the coefficient of the ``$1$'' term in the 
expansion of $\sqrt{-g}$.

\begin{figure}
\centering
\epsfig{file=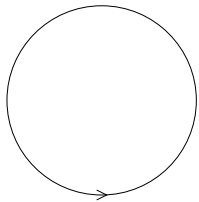,width= .15\linewidth}
\caption{Free particle vacuum energy diagram in SGET.}
\end{figure} 

\begin{figure}
\centering
\epsfig{file=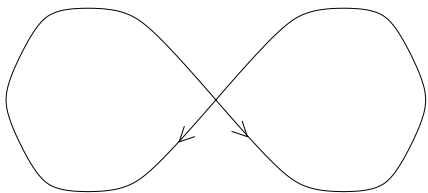,width= .33 \linewidth}
\caption{A vacuum diagram with propagators from a vertex to itself.}
\end{figure}

Let us consider some typical diagrams. We have already discussed the non-interacting 
diagram of Fig. 15 in the 
previous section and explained why it must be set to zero. Fig. 16 is an example of 
a vacuum diagram involving propagators from a vertex to itself, which apparently 
requires us to make sense of expressions such as 
\begin{equation}
\frac{\sum_{v} \int d^4 k}{RPI} ~ \frac{1}{k.v + i \epsilon}.
\end{equation}
However, 
as was the case in Fig. 15, there is a physical reason why we must take the diagram to 
vanish. The reason is that Fig. 16 requires a vertex of the form 
$\phi_{v_1} \phi_{v_2} ~\phi^{\dagger}_{v_1} \phi^{\dagger}_{v_2}$, obtained by 
matching in situations which there is only a soft momentum transfer in two-particle 
scattering. As in the discussion of Fig. 15, under these circumstances the effective 
theory cannot distinguish whether the heavy particle is itself a composite of very light 
particles or solitonic,  in which case it cannot make robust large loop 
contributions. We can summarize the vanishing of diagrams like Fig. 15 and 16 by 
saying that in the SGET we  must normal order.

\begin{figure}[t]
\centering
\epsfig{file=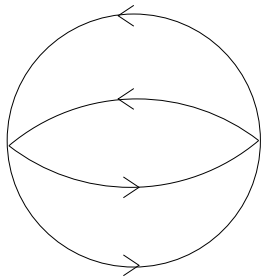,width=.2\linewidth}
\caption{A vacuum diagram in SGET which survives normal ordering, but not evaluation.
}
\end{figure}

Fig. 17 is a vacuum diagram one can draw despite normal ordering. We will compute 
it by first gauge fixing in a similar manner to the 
non-vacuum loop diagram we examined above. However, unlike that case
 where there was a natural frame selected by the incoming (or outgoing) momenta, we 
must simply choose a frame arbitrarily, and fix the gauge $\vec{k}_{int} = 0$ in that 
frame. One might worry that this will lead to a Lorentz non-invariant answer, but it will
not as we will see below. There are originally four $k^0_{int}$ residual energies for 
the four internal lines, but after integrating the energy-conserving $\delta$-function
we are left with three residual energy integrals to be done. Let us focus on any 
single one of these. It clearly has the form
\begin{eqnarray}  
\int d k^0_{int} ~ \frac{1}{k^0_{int} v^0_{int} + i \epsilon} ~ 
\frac{1}{k^0_{int} v'^{0}_{int} + ... + i \epsilon} ~ = ~ 0,
\end{eqnarray}
where the ellipsis refers to a (real) combination of other energies external to this 
integral. This integral is finite and evaluates to zero by contour integration. To see 
this note that $v^0_{int},  v'^{0}_{int} > 0$ and therefore
both poles lie on the same side of the real $k^0_{int}$ axis, so the contour 
can be closed at infinity without enclosing any poles.

\begin{figure}[t]
\centering
\epsfig{file=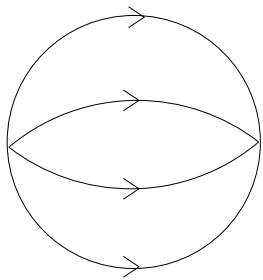,width=.2\linewidth}
\caption{A non-existent vacuum diagram in SGET because of the absence of the 
requisite vertices.}
\end{figure}

If we repeated this same type of analysis for Fig. 18, we would get residual energy 
integrals of the form
\begin{eqnarray}  
\int d k^0_{int} ~ \frac{1}{k^0_{int} v^0_{int} + i \epsilon} ~~ 
\frac{1}{- k^0_{int} v'^{0}_{int} + ... + i \epsilon},
\end{eqnarray}
which does not vanish. However, the diagram does not exist in the SGET because it 
requires vertices of the form $\phi \phi \phi \phi$ rather than the
$\phi^{\dagger} \phi^{\dagger} \phi \phi$ vertices appearing in Fig. 17. 
Recalling that SGET vertices arise from matching to fundamental 
correlators with nearly on-shell external lines, we see that no effective vertex such as 
$\phi \phi \phi \phi$ could have arisen upon matching. Therefore Fig. 18 simply does not 
exist. Recalling that $\phi^{\dagger}_v$ and $\phi_v$ are creation and destruction 
operators, one might wonder whether the presence of 
$\phi^{\dagger} \phi^{\dagger} \phi \phi$ is correlated with that of 
$\phi \phi \phi \phi$ as it is in fundamental field theories. In fundamental theories 
this is a consequence of locality, there is no local way of separating positive and 
negative energy operators. However, as we have seen above, locality only down to 
$\Lambda_{grav} \ll m$ does not imply such a correlation. 

It is straightforward to check that the examples of Figs. 15 to 18 exhaust all the 
possible cases arising in general vacuum diagrams. In SGET there are simply no robust 
contributions from heavy matter to the cosmological constant! 

For a fat graviton there is 
new gravitational physics at $\Lambda_{grav}$, its
vibrational excitations. We do not explicitly know this physics but 
we can estimate its loop contributions to the cosmological constant by standard 
power-counting, ${\cal O}(\Lambda_{grav}^4/16 \pi^2)$. This sets the minimal natural 
size of the cosmological constant. 

\section{Conclusions}

The soft graviton effective theory demonstrates a clear qualitative 
distinction between (a)
loop effects of heavy SM physics on  SM processes, (b) soft graviton exchanges 
between such ongoing SM processes, and (c) loop effects of heavy SM physics on
the low-energy 
gravitational effective action, and in particular the infrared cosmological 
constant. The effective theory can match (a) to the fundamental SM theory (which 
is of course very well tested), and describe (b) constrained only by 
general coordinate invariance in the infrared, such as even a fat graviton must have. 
In this manner the effective theory captures the two pillars of our experimental 
knowledge, soft gravitation and the SM of high-energy physics.
However, none of this implies any robust contributions to (c). 

There is therefore a loop-hole 
in the cosmological constant problem for a fat graviton which is absent for a point-like 
graviton, and it makes sense to 
vigorously hunt for its realization experimentally  and 
within more fundamental theories such as string theory.
Quantum naturalness related to the contributions to the cosmological constant 
of the vibrational excitations of 
the fat graviton, as well as from soft gravitons, photons and neutrinos,
 implies that Newton's Law should yield to a suppression of 
the gravitational force below distances of {\it roughly} 20 microns, as illustrated 
in Fig. 8. Of course the onset of such modifications of Newton's Law may be seen at somewhat 
larger distances.

Future work will focus on generalizing the effective theory to include 
massless or light SM particles as well as quantum gravity corrections, and to 
looking for potential signals for experiment and observation 
due to higher order (in the derivative expansion) effects,
that are allowed by the effective theory but forbidden by the standard  
theory where point-like gravity is extrapolated to at least a TeV. 
The compatibility of fat graviton ideas with multiple matter vacua will also be investigated.

\section*{Acknowledgements}

In thinking about fat gravitons,
I greatly benefitted from comments, criticisms, leads and help from  Nima Arkani-Hamed,
 Tom Banks, Andy Cohen, Savas Dimopoulos, Michael Douglas, Gia Dvali, Adam Falk,
Shamit Kachru,  Mike Luke, Markus Luty, Ann Nelson, Dan Pirjol, Joe Polchinski, Martin 
Schmaltz, Matt Strassler and Andy Strominger. 
This research was supported in part by 
NSF Grant P420D3620434350 and in part by 
DOE Outstanding Junior Investigator Award Grant P442D3620444350.

\end{document}